\documentclass{PoS}

\newcommand{\be}{\begin{equation}}
\newcommand{\ee}{\end{equation}}
\newcommand{\bea}{\begin{eqnarray}}
\newcommand{\eea}{\end{eqnarray}}
\newcommand{\non}{\nonumber}

\title{Update on $N_{\rm f}=3$ finite temperature QCD phase structure with Wilson-Clover fermion action}

\ShortTitle{Update on $N_{\rm f}=3$ finite temperature QCD phase structure with Wilson-Clover fermion action}

\author{
   \speaker{Shinji Takeda}$^{a,b}$\thanks{
This research used computational resources of 
HA-PACS and COMA provided by Interdisciplinary Computational Science Program in Center for Computational Sciences, University of Tsukuba,
System E at Kyoto University through the HPCI System Research project (Project ID:hp150141) and
PRIMERGY CX400 tatara at Kyushu University.
This work is supported by JSPS KAKENHI Grant Numbers 26800130.
This work was supported by FOCUS Establishing Supercomputing Center of Excellence
and Kanazawa University SAKIGAKE Project.
This research used resources of the Argonne Leadership Computing Facility, which is a DOE Office of Science User Facility supported under Contract DE-AC02-06CH11357.
   }, 
   Xiao-Yong Jin$^c$,
   Yoshinobu Kuramashi$^{b,d,e}$,
   Yoshifumi Nakamura$^{b,f}$,
   Akira Ukawa$^b$
   \\
   E-mail: takeda@hep.s.kanazawa-u.ac.jp
   \\
   \\
   \llap{$^a$}
Institute of Physics, Kanazawa University, Kanazawa 920-1192, Japan
   \\
   \llap{$^b$}
RIKEN Advanced Institute for Computational Science,
Kobe, Hyogo 650-0047, Japan
   \\
   \llap{$^c$}
Argonne Leadership Computing Facility, Argonne National Laboratory, Argonne, IL 60439, USA
   \\
   \llap{$^d$}
Graduate School of Pure and Applied Sciences,
University of Tsukuba,
Tsukuba, Ibaraki 305-8571, Japan
   \\
   \llap{$^e$}
Center for Computational Sciences,
University of Tsukuba,
Tsukuba, Ibaraki 305-8577, Japan
   \\
   \llap{$^f$}
Graduate School of System Informatics, Department of Computational Sciences, Kobe University, Kobe, Hyogo 657-8501, Japan
}

\abstract{
We present an update of the finite temperature phase structure analysis for three flavor QCD.
In the study the Iwasaki gauge action and non-perturvatively O($a$) improved Wilson-Clover fermion action
are employed.
We discuss finite size scaling analysis including mixings of magnetization-like and energy-like observables. 
Preliminary results are shown  of the continuum limit of the critical point using newly generated data
at $N_{\rm t}=8,10$, including estimates of  the critical pseudo-scalar meson mass and critical temperature.
}

\FullConference{34th annual International Symposium on Lattice Field Theory\\
		 24-30 July 2016\\
		 University of Southampton, UK}

\begin{document}

\section{Introduction}
The nature of finite temperature transition in QCD varies depending on the quark masses.
A pictorial representation is often given as the Columbia plot whose axes
are usually taken to be  the up-down and strange quarks mass.
In this report we restrict ourselves to the case of zero quark number density.

There are two longstanding issues on the plot, namely
the location of the critical line which separates the first  order phase transition region from the cross over region, and the universality class of the critical line. 
Studies with  the standard staggered fermion action \cite{Kaya99,Karsch01,Philipsen07,Smith11}
successfully located the critical point along the flavor symmetric line ($N_{\rm f}=3$).  
It was subsequently found that the first order region rapidly shrinks towards the continuum limit \cite{deForcrand:2007rq}.
Further studies with  staggered fermions with smearing techniques \cite{Endrodi07,Ding11}
could not even detect a critical point, perhaps due to the possibility that the critical quark mass is so small that current computational resources cannot access it.

On the other hand, the pioneering Wilson-type fermion study in Ref.~\cite{Iwasaki96} reported a relatively heavy critical mass.
Our recent study \cite{cep3ft}, while confirming a large value for coarse lattice spacings, suggested that the critical mass appears to be smaller for finer lattice spacings.
This implies  that removal of  scaling violation is crucial for Wilson-type fermion action as well.
In Ref.~\cite{cep3ft}, we computed the critical point for $N_{\rm f}=3$ QCD at temporal lattice sizes $N_{\rm t}=4$, $6$ and $8$.
In order to take the continuum limit more reliably, we have recently started large scale simulations at $N_{\rm t}=10$
and a preliminary result was already reported in the previous lattice conference \cite{nakamuralattice2015}.
In this manuscript, we update the data at $N_{\rm t}=8,10$ and
examine the continuum limit with the added data. 

Concerning the issue with the universality class along the critical line,  we observed in Ref.~\cite{cep3ft,nakamuralattice2015} that for $N_{\rm t}=8$ and $10$ the value of kurtosis at the critical point  
is somewhat different from that of three-dimensional Z$_2$ universality class,
in contrast to the situation with $N_{\rm t}=4$ and $6$ where it is consistent.  
We address this issue by noting that bare lattice observables generally are  mixtures of magnetization-like and energy-like operators, which should be taken into account in finite size scaling analyses.

\section{Setup and methods}
We employ the Iwasaki gauge action \cite{iwasaki} and non-perturvatively O($a$) improved Wilson-Clover fermion action \cite{csw}
to carry out the finite temperature $N_{\rm f}=3$ QCD simulation.
The temporal lattice size we newly report here is $N_{\rm t}=8$ and $10$.
The spatial lattice size is varied over $N_{\rm s}=16$, $20$, $24$ and $28$ to carry out  finite size scaling.
BQCD code \cite{BQCD} implementing the RHMC algorithm is used to generate gauge configurations, with the acceptance rate tuned to be around 80\%. 
We store configurations at every 10th trajectory for observable measurements. 
For each parameter set, we accumulate O($1,000-10,000$) configurations.
Since the three dynamical quarks are all degenerate, we have only one hopping parameter $\kappa$.
Several values  of the parameter $\beta$ (3 to 4) are selected,  and  $\kappa$ is adjusted
to search for a transition point at each $\beta$.

We use the naive chiral condensate as a probe to study the phase structure.
We also measure higher moments of the chiral condensate up to 4th order to calculate 
susceptibility, skewness and kurtosis.  
In order to determine the transition point, 
we use the susceptibility peak position,  and confirm it with the skewness zero.
The kurtosis is used to locate the critical point through the  intersection analysis. 

We combine several ensembles, which share common parameter values except for $\kappa$,
by the multi-ensemble reweighting in $\kappa$ for interpolating the moments. We do not apply $\beta$-reweighting.  
To calculate the reweighting factor given by the ratio of fermion determinants
at different $\kappa$ values,
we use an  expansion of the logarithm of the determinant \cite{Kuramashi:2016kpb}.
For the computation of the observable part in the reweighting procedure,
we need to evaluate quark propagators at continuously many points of $\kappa$.
We adopt an expansion form for the moments which allows us to evaluate the moments at continuously many points at a relatively low cost.

\section{Scaling analysis for general observable}

In our scaling analysis,
the relevant parameters are reduced temperature $t$, external magnetic field $h$ and the inverse of the linear lattice size $L^{-1}$.
According to finite size scaling theory, under scaling by a factor $b$, 
the free energy (not free energy density) scales as follows up to analytic terms:
\be
F(t,h,L^{-1})=F(tb^{y_{\rm t}},hb^{y_{\rm h}},L^{-1}b),
\label{eqn:Frescale}
\ee
where $y_{\rm t}$ and $y_{\rm h}$ are the temperature and the magnetic exponent, respectively.
Setting $b=L$, 
the scaling relation of the free energy is given by
\be
F(t,h,L^{-1})=F(tL^{y_{\rm t}},hL^{y_{\rm h}},1).
\ee
We use the following notation and abbreviation in the following:
\be
F(tL^{y_{\rm t}},0,1)
=
F(tL^{y_{\rm t}}),
\hspace{10mm}
\frac{\partial^n}{\partial t^{n}}
\frac{\partial^m}{\partial h^{m}}
F(t,h,L^{-1})
=
F^{(nm)}(t,h,L^{-1}).
\ee
%

The kurtosis for magnetization operator  ${\cal M}$ at $h=0$ is then given by
\be
K(t,0,L^{-1})
=
\frac
{\left.F^{(04)}(t,h,L^{-1})\right|_{h=0}}
{\left[\left.F^{(02)}(t,h,L^{-1})\right|_{h=0}\right]^2}
=
\frac
{L^{4y_{\rm h}}F^{(04)}(tL^{y_{\rm t}})}
{\left[L^{2y_{\rm h}}F^{(02)}(tL^{y_{\rm t}})\right]^2}
=
\frac
{F^{(04)}(tL^{y_{\rm t}})}
{\left[F^{(02)}(tL^{y_{\rm t}})\right]^2}.
\ee
At $t=0$, the kurtosis is independent of the volume.
For small $tL^{y_{\rm t}}$,
one can expand
\be
K(t,0,L^{-1})
=
\frac
{F^{(04)}(0)}
{\left[F^{(02)}(0)\right]^2}
+
c_K
tL^{1/\nu}
+...
\ee
where we have used $y_{\rm t}=1/\nu$.
This is the well known formula for the kurtosis intersection analysis.

For a general observable ${\cal O}$ which is a mixture of energy ${\cal E}$ and magnetization ${\cal M}$,
\be
{\cal O}=
c_{\rm M}
{\cal M}
+
c_{\rm E}
{\cal E}
\rightarrow
c_{\rm M}
\frac{\partial}{\partial h}
+
c_{\rm E}
\frac{\partial}{\partial t},
\ee
the kurtosis for ${\cal O}$ at $h=0$
is given by,
\bea
&&
K_{\cal O}(t,0,L^{-1})
=
\frac{
\left.
\left(
c_{\rm M}
\frac{\partial}{\partial h}
+
c_{\rm E}
\frac{\partial}{\partial t}
\right)^4
F(t,h,L^{-1})
\right|_{h=0}
}{
\left[
\left.
\left(
c_{\rm M}
\frac{\partial}{\partial h}
+
c_{\rm E}
\frac{\partial}{\partial t}
\right)^2
F(t,h,L^{-1})
\right|_{h=0}
\right]^2
}
\non\\
&&=
\frac{F^{(04)}(tL^{y_{\rm t}})}{F^{(02)}(tL^{y_{\rm t}})^2}
\left[
1+
\frac{4c_{\rm E}}{c_{\rm M}}
L^{y_{\rm t}-y_{\rm h}}
\left(
\frac{F^{(13)}(tL^{y_{\rm t}})}{F^{(04)}(tL^{y_{\rm t}})}
-
\frac{F^{(11)}(tL^{y_{\rm t}})}{F^{(02)}(tL^{y_{\rm t}})}
\right)
+
O(L^{2(y_{\rm t}-y_{\rm h})})
\right].
\label{eqn:kurtosisO}
\eea
Thus, even when setting $t=0$,
the correction term of $O(c_{\rm E}L^{y_{\rm t}-y_{\rm h}})$ alters the value of kurtosis at the critical point.
The difference of the exponents $y_{\rm t}-y_{\rm h}$ is usually negative for various universality classes, {\it viz.}
\be
y_{\rm t}-y_{\rm h}
=
\frac{1}{2\nu}(\alpha-\gamma)
=
\left\{
\begin{array}{lll}
\frac{1}{2\cdot1}(0-7/4)&=-7/8 &\mbox{ : 2D Ising},\\
\frac{1}{2\cdot0.630}(0.110-1.237)&=-0.894 &\mbox{ : 3D Ising},\\
\frac{1}{2\cdot0.67}(-0.01-1.32)&=-0.993 &\mbox{ : 3D O(2)},\\
\frac{1}{2\cdot0.75}(-0.25-1.47)&=-1.15 &\mbox{ : 3D O(4)}.
\end{array}
\right.
\ee
Therefore, such a correction would be irrelevant in the large volume limit.  
However, at finite volumes,  the value of kurtosis at $t=0$ has a volume dependence;  
the kurtosis for various volumes would not cross at a single point. 

\section{Results}
\subsection{Kurtosis analysis}

\begin{figure}[t!]
\begin{center}
\begin{tabular}{cc}
\scalebox{0.63}{\includegraphics{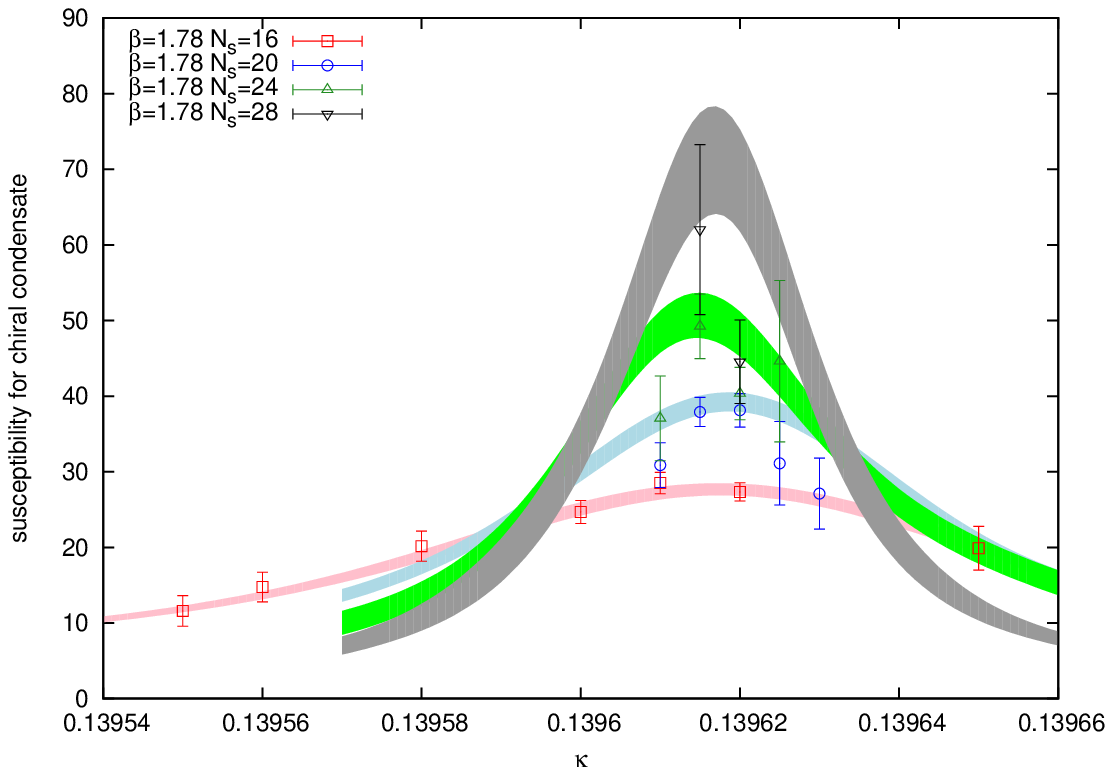}}
&
\scalebox{0.63}{\includegraphics{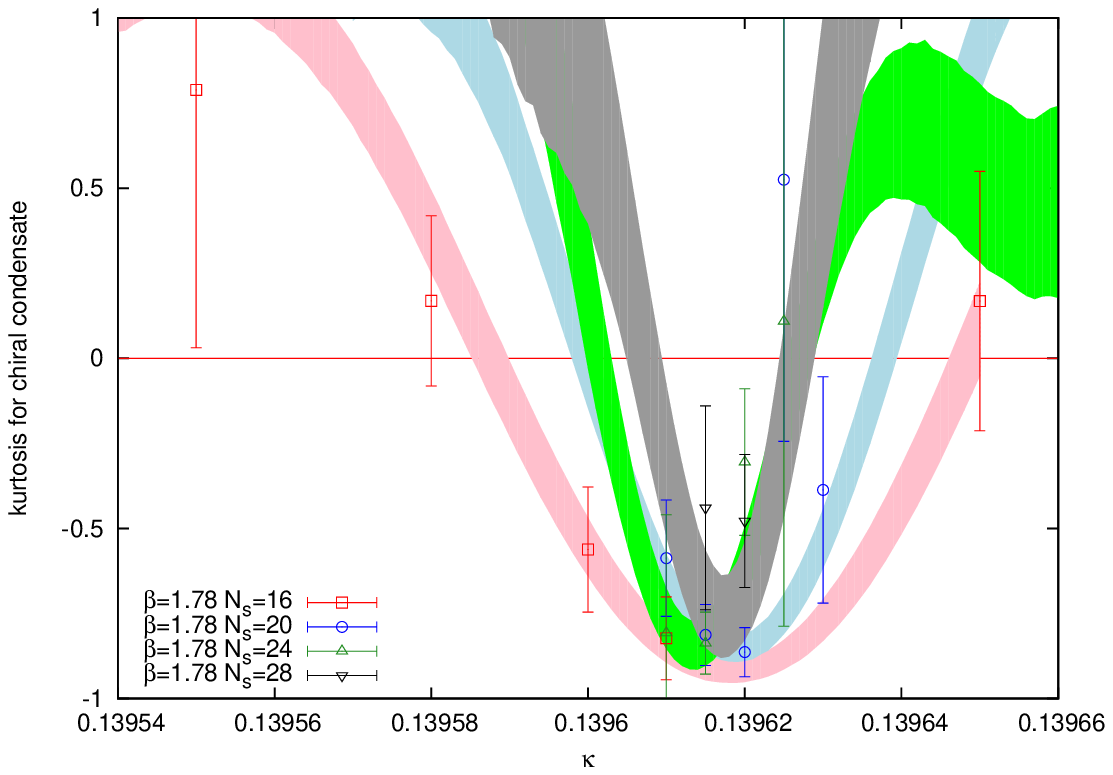}}
\\
\end{tabular}
\end{center}
\caption{The susceptibility (left) and kurtosis (right) of chiral condensate as a function $\kappa$ for $(N_{\rm t},\beta)=(10,1.78)$ with several spatial sizes, $N_{\rm s}=16-28$.
The raw data points as well as the multi-ensemble reweighting (1-$\sigma$ band) are plotted.
}
\label{fig:moment}
\end{figure}

\begin{figure}[h!]
\begin{center}
\begin{tabular}{c}
\scalebox{0.75}{\includegraphics{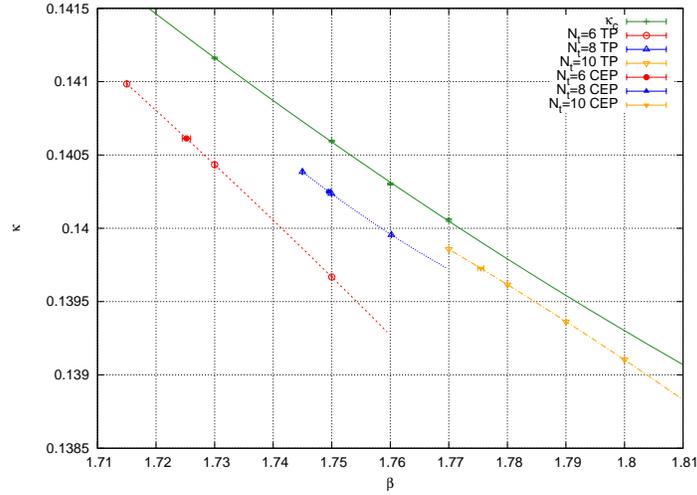}}
\end{tabular}
\end{center}
\vspace{-3mm}
\caption{
Phase diagram for bare parameter space $(\beta,\kappa)$ at $N_{\rm t}=6$, $8$ and $10$.
The open symbols represent a transition point (TP) while the filled symbols are critical point (CEP) which
is determined by the kurtosis intersection using new formula.
$\kappa_{\rm c}$ is the pseudo-scalar massless point with $N_{\rm f}=3$ at the zero temperature.
}
\label{fig:phase_diagram}
\end{figure}

\begin{figure}[h!]
\begin{center}
\begin{tabular}{c}
\scalebox{1.1}{\includegraphics{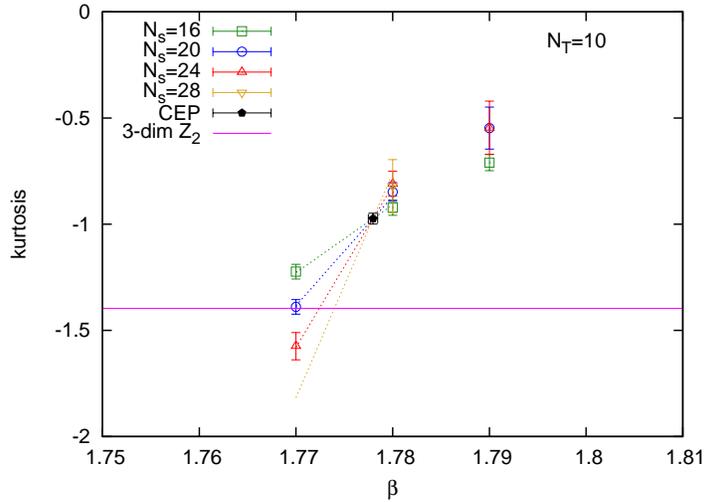}}
\end{tabular}
\end{center}
\vspace{-6mm}
\caption{
Kurtosis intersection for chiral condensate with the simple fitting form: $K=K_{\rm E}+AN_{\rm s}^{1/\nu}(\beta-\beta_{\rm E})$.
Obviously the value of kurtosis at the crossing point (black pentagon) is not consistent with the three-dimensional Z$_2$ universality class.
}
\label{fig:krt}
\end{figure}

As an illustration of new data, we show the susceptibility and the kurtosis of the chiral condensate
for $(N_{\rm t},\beta)=(10,1.78)$ in Fig.~\ref{fig:moment}
together with the $\kappa$-reweighting results.
From the peak position of the susceptibility, we extract the transition points
and the phase diagram of the bare parameter space as summarized in Fig.~\ref{fig:phase_diagram}.

The minimum of kurtosis at each $(N_{\rm s},\beta)$ is plotted in Fig.~\ref{fig:krt}
to perform  kurtosis intersection analysis at $N_{\rm t}=10$.
Clearly an analysis using the conventional formula given in the caption of Fig.~\ref{fig:krt} 
leads to a value for $K_{\rm E}$ which is  substantially larger than that for the 3D Z$_2$ universality class.
%
%
In this situation we attempt a modified fitting form  which
incorporates the correction term in (\ref{eqn:kurtosisO})
associated with the contribution of the energy-like observable given by 
\be
K
=
\left[
K_{\rm E}
+
AN_{\rm s}^{1/\nu}(\beta-\beta_{\rm E})
\right]
(1+BN_{\rm s}^{y_{\rm t}-y_{\rm h}}),
\label{eqn:new}
\ee
where we have two additional parameters $B$ and $y_{\rm t}-y_{\rm h}$.
The fit results are given in Table \ref{tab:CEP}  with a description of "Fit-1, 2, 3" in the caption.

For $N_{\rm t}=6$, the parameter $B$ in "Fit-3" is consistent with zero
and all other fitting parameters of all fitting forms are consistent with each other.
Thus we conclude that the new correction term is negligible and universality class is  consistent with 3D Z$_2$ for $N_{\rm t}=6$.

For $N_{\rm t}=8$ and $10$, 
the assumption of Z$_2$ universality class is unlikely to hold without the new correction term 
since the Fit-2, which assumes the $Z_2$ values for $K_{\rm E}$ and $\nu$,  has  large $\chi^2/{\rm d.o.f.}$.  
On the other hand, with the Fit-3 assuming $Z_2$ but including the correction term from mixing of magnetization and energy terms, we observe a reasonable $\chi^2/{\rm d.o.f.}<1$.  The magnitude of the correction term is reasonably small of order 10\%.  
This suggests that $N_{\rm t}=8,10$ results are  consistent with the 3D Z$_2$ universality class
if one includes the  correction term.
In future, to enhance reliability of our analysis, we shall
perform the mixed observable analysis which eliminates the energy part before doing the kurtosis intersection analysis.

\begin{table}[t!]
\begin{center}
{\scriptsize
\begin{tabular}{rrllllllrr}
\hline\hline
Fit
&
$N_{\rm t}$
&
$\beta_{\rm E}$
&
$\kappa_{\rm E}$
&
$K_{\rm E}$
&
$\nu$
&
$A$
&
$B$
&
$y_{\rm t}-y_{\rm h}$
&
$\chi^2/{\rm d.o.f.}$
\\
\hline
1&
$6$&$1.72518 ( 71 )$&$0.1406129 ( 14 )$&$-1.373 ( 17 )$&$0.683 ( 54 )$&$0.58 ( 17 )$&$\times$&$\times$&$0.68$\\
2&
$6$&$1.72431 ( 24 )$&$0.1406451 ( 14 )$&$-1.396$&$0.63$&$0.418 ( 11 )$&$\times$&$\times$&$0.70$\\
3&
$6$&$1.72462 ( 40 )$&$0.1406334 ( 14 )$&$-1.396$&$0.63$&$0.422 ( 12 )$&$-0.052 ( 52 )$&$-0.894$&$0.70$\\
\hline
1&
$8$&$1.75067 ( 40 )$&$0.1402147 ( 16 )$&$-1.211 ( 16 )$&$0.534 ( 37 )$&$0.155 ( 62 )$&$\times$&$\times$&$0.29$\\
2&
$8$&$1.74686 ( 52 )$&$0.14032912 ( 79 )$&$-1.396$&$0.63$&$0.388 ( 36 )$&$\times$&$\times$&$4.99$\\
3&
$8$&$1.74953 ( 29 )$&$0.1402488 ( 13 )$&$-1.396$&$0.63$&$0.408 ( 11 )$&$-1.34 ( 12 )$&$-0.894$&$0.38$\\
\hline
1&
$10$&$1.77796 ( 48 )$&$0.1396661 ( 17 )$&$-0.974 ( 25 )$&$0.466 ( 45 )$&$0.084 ( 52 )$&$\times$&$\times$&$0.22$\\
2&
$10$&$1.7694 ( 16 )$&$0.1398724 ( 22 )$&$-1.396$&$0.63$&$0.421 ( 95 )$&$\times$&$\times$&$10.03$\\
3&
$10$&$1.77545 ( 53 )$&$0.1397274 ( 17 )$&$-1.396$&$0.63$&$0.559 ( 29 )$&$-2.97 ( 25 )$&$-0.894$&$0.43$\\
\hline\hline
\end{tabular} }
\end{center}
\caption{
Fit results for kurtosis intersection with fitting form in eq.(4.1).
"Fit 1": no correction ($B$) term and all parameters are used as fit parameter.
"Fit 2": no correction term and assuming the 3D Z$_2$ universality class for $K_{\rm E}$ and $\nu$.
"Fit 3": including correction term and assuming the 3D Z$_2$ universality class for $K_{\rm E}$, $\nu$ and $y_{\rm t}-y_{\rm h}$.
A value without error bar means that the corresponding fit parameter is fixed to the written value during the fit.
For the 3D Z$_2$ universality class, the expected values of the parameter are
$K_{\rm E}=-1.396$, $\nu=0.630$ and $y_{\rm t}-y_{\rm h}=-0.894$ respectively.
}
\label{tab:CEP}
\end{table}

\subsection{Continuum extrapolation of critical pseudo scalar meson  mass and critical temperature}

\begin{figure}[t]
\begin{center}
\begin{tabular}{c}
\scalebox{0.8}{\includegraphics{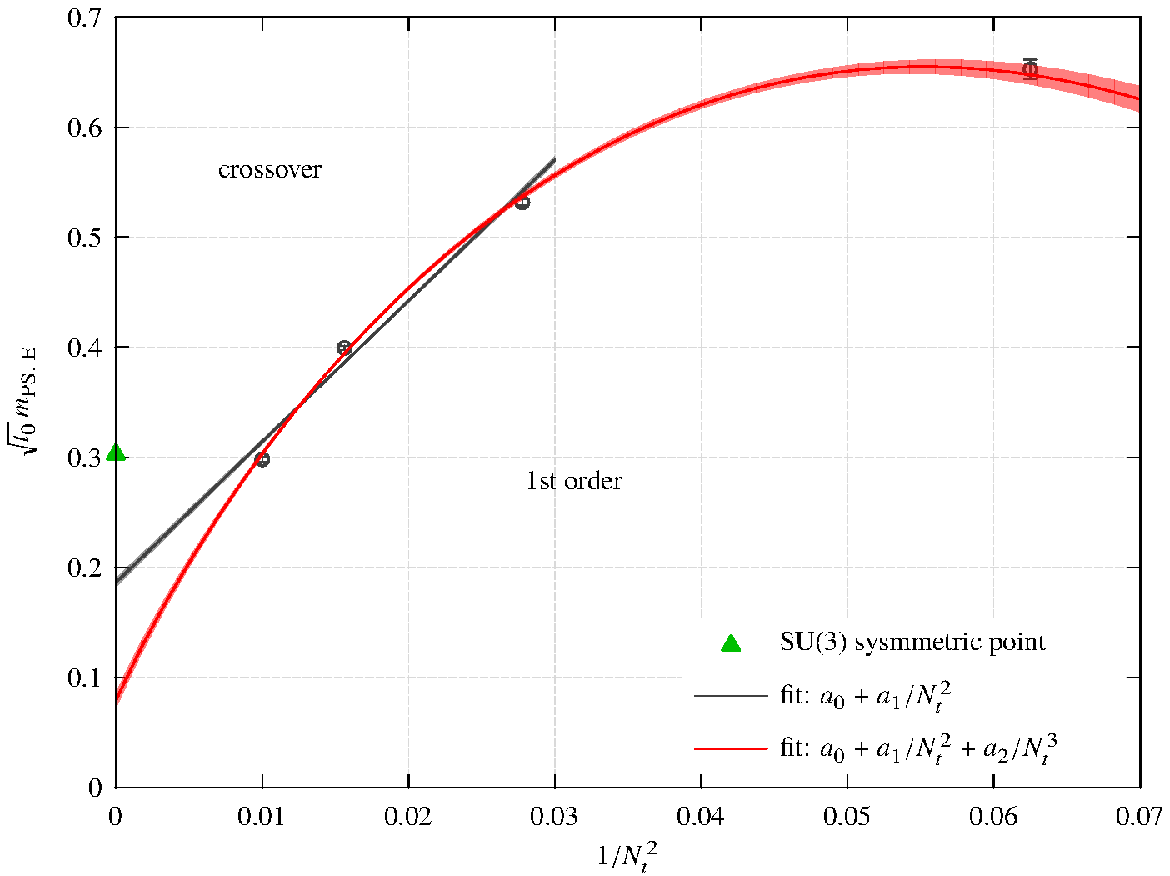}}
\\
\scalebox{0.8}{\includegraphics{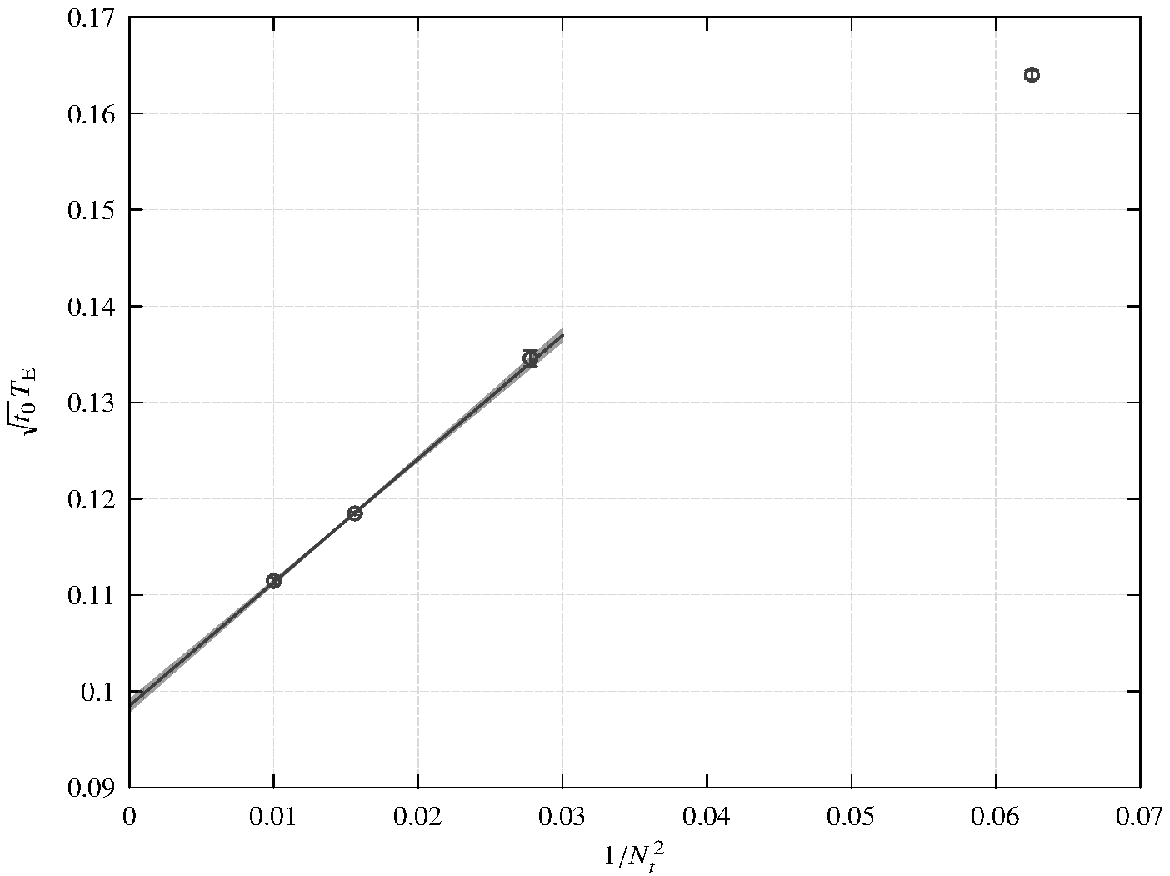}}
\end{tabular}
\end{center}
\caption{
Continuum extrapolation of the critical point
$\sqrt{t_0}m_{\rm PS,E}$ (upper) and $\sqrt{t_0}T_{\rm E}$ (lower)
normalized by the Wilson flow scale $\sqrt{t_0}$.
}
\label{fig:continuum}
\end{figure}

Finally, in Fig.~\ref{fig:continuum} we show the continuum extrapolation of the critical pseudo scalar meson mass  
$\sqrt{t_0}m_{\rm PS,E}$ and critical temperature $\sqrt{t_0}T_{\rm E}$ normalized by the Wilson flow scale $\sqrt{t_0}$ \cite{wilflow}.
To gain some idea of the critical point in physical units, we carry out continuum extrapolation including up to cubic term in the lattice spacing and set the Wilson flow scale $1/\sqrt{t_0}=1.347(30)$GeV
\cite{t0BMW}. 
We then  obtain
$m_{\rm PS,E}
\approx
100 {\rm MeV}.
$
This value is much smaller than our previous estimate ($\sim$300MeV) \cite{cep3ft}, the reason being 
that the latest point at $N_{\rm t}=10$ (upperl panel of Fig.~\ref{fig:continuum}) bends down the continuum extrapolation. 
Since the scaling violation is quite large, universality check of the above value
by using a different lattice action would be essential.

\end{document}